\def\papertitle{Physical Modeling using Recurrent Neural Networks with Fast Convolutional Layers}
\def\paperauthorA{Julian D. Parker}
\def\paperauthorB{Sebastian J. Schlecht}
\def\paperauthorC{Rudolf Rabenstein}
\def\paperauthorD{Maximilian Schäfer}

\documentclass[twoside,a4paper]{article}
\usepackage{etoolbox}
\usepackage{dafx_22a}
\usepackage{amsmath,amssymb,amsfonts,amsthm}
\usepackage{bm}
\usepackage{mathtools}
\usepackage{siunitx}
\usepackage{euscript}
\usepackage[T1]{fontenc}
\usepackage[utf8]{inputenc}
\usepackage{nimbusserif}
\usepackage{ifpdf}
\usepackage[english]{babel}
\usepackage{caption}
\usepackage{subcaption} 
\usepackage{xcolor}
\usepackage{tikz}
\usepackage{tikzscale}
\usepackage{cite}
\usepackage{trfsigns}
\usepackage{url}
\usepackage{lipsum}
\usepackage{booktabs}

\input glyphtounicode
\ninept

\newcounter{numauth}
\newcounter{listcnt}
\newcommand\authcnt[1]{\ifdefined#1 \stepcounter{numauth} \fi}

\newcommand\addauth[1]{
\ifdefined#1 
\stepcounter{listcnt}
\ifnum \value{listcnt}<\value{numauth}
\appto\authorslist{, #1}
\else
\appto\authorslist{~and~#1}
\fi
\fi}
\authcnt{\paperauthorB}
\authcnt{\paperauthorC}
\authcnt{\paperauthorD}
\authcnt{\paperauthorE}
\authcnt{\paperauthorF}
\authcnt{\paperauthorG}
\authcnt{\paperauthorH}
\authcnt{\paperauthorI}
\authcnt{\paperauthorJ}
\def\authorslist{\paperauthorA}
\addauth{\paperauthorB}
\addauth{\paperauthorC}
\addauth{\paperauthorD}
\addauth{\paperauthorE}
\addauth{\paperauthorF}
\addauth{\paperauthorG}
\addauth{\paperauthorH}
\addauth{\paperauthorI}
\addauth{\paperauthorJ}

\usepackage{times}

\newif\ifpdf
\ifx\pdfoutput\relax
\else
   \ifcase\pdfoutput
      \pdffalse
   \else
      \pdftrue
\fi

\ifpdf 
  \usepackage[pdftex,
    pdftitle={\papertitle},
    pdfauthor={\authorslist},
    pdfsubject={Proceedings of the 25th International Conference on Digital Audio Effects (DAFx20in22)},
    colorlinks=false, 
    bookmarksnumbered, 
    pdfstartview=XYZ 
  ]{hyperref}
\else 
\fi

\usepackage[hypcap=true]{caption}
\title{\papertitle}

\fouraffiliations{
\paperauthorA\sthanks{Equal contribution}}
{\href{https://www.native-instruments.com/en/}{Native Instruments GmbH} \\ Berlin, Germany\\
{\tt \href{mailto:firstname.lastname@native-instruments.de}{firstname.lastname@native-instruments.de}}
}
{\paperauthorB\footnotemark[1]}
{\href{}{Aalto Acoustics Lab and Media Lab} \\ Aalto University \\ Espoo, Finland \\ {\tt \href{mailto:sebastian.schlecht@aalto.fi}{sebastian.schlecht@aalto.fi}}
}
{\paperauthorC\footnotemark[2]\footnotemark[1]}
{\href{https://www.lms.tf.fau.eu/}{Multimedia Communications and Signal Processing} \\ Friedrich-Alexander-Universit{\"a}t Erlangen-N{\"ur}nberg (FAU)\\ Erlangen, Germany\\ {\tt \href{mailto:Rudolf.Rabenstein@fau.de}{Rudolf.Rabenstein@fau.de}}
}
{\paperauthorD\sthanks{This work was supported by the German Research Foundation (DFG) under grant number RA 807/7-1}\footnotemark[1]}
{\href{https://www.idc.tf.fau.de}{Digital Communications} \\ Friedrich-Alexander-Universität Erlangen-Nürnberg (FAU) \\ Erlangen, Germany \\ {\tt \href{mailto:dafx2019@gmail.com}{max.schaefer@fau.de}}
}
\newcommand{\dint}[1]{\,\mathrm{d}#1}
\newcommand{\Kprim}{\bm{K}}
\newcommand{\Kadj}{\tilde{\bm{K}}}
\newcommand{\her}{^{\scriptscriptstyle\mathrm{H}}}
\newcommand{\tran}{^{\scriptscriptstyle\mathrm{T}}}

\newcommand{\As}{\bm{\mathcal{A}}}
\newcommand{\Cs}{\bm{\mathcal{C}}}
\newcommand{\Csa}{\tilde{\bm{\mathcal{C}}}}

\newcommand{\xn}{\mathbf{x}_\mathbf{n}}

\begin{document}
\ifpdf 
  \DeclareGraphicsExtensions{.png,.jpg,.pdf}
\else  
  \DeclareGraphicsExtensions{.eps}
\fi


\maketitle

\begin{abstract}
\sloppy
Discrete-time modeling of acoustic, mechanical and electrical systems is a prominent topic in the musical signal processing literature. Such models are mostly derived by discretizing a mathematical model, given in terms of ordinary or partial differential equations, using established techniques. Recent work has applied the techniques of machine-learning to construct such models automatically from data for the case of systems which have lumped states described by scalar values, such as electrical circuits. In this work, we examine how similar techniques are able to construct models of systems which have spatially distributed rather than lumped states. We describe several novel recurrent neural network structures, and show how they can be thought of as an extension of modal techniques. As a proof of concept, we generate synthetic data for three physical systems and show that the proposed network structures can be trained with this data to reproduce the behavior of these systems.
\fussy
\end{abstract}

\section{Introduction}
\label{sec:intro}

Discrete-time modeling of systems (both acoustic, mechanical \& electrical) which are relevant to musical uses has a long history in the literature. This discipline is known as \emph{physical modeling}, or as \emph{virtual-analog} when referring to the subset of electrical systems. Popular approaches to such problems in the acoustic or mechanical domain include direct numerical solution of \emph{ordinary differential equations} (ODEs) or \emph{partial differential equations} (PDEs) via finite differences \cite{Bilbao:2009}, digital waveguides \cite{Bensa:JASA:2003} and modal approaches like the Functional Transformation Method (FTM) \cite{Trautmann2003DigitalMethod,Schafer2019SimulationModels}. Popular approaches in the electrical domain include finite differences or state-space models \cite{Holters2015StateSpace}, wave digital filters \cite{fettweis:ieee:1986} and the Port-Hamiltonian formalism \cite{Falaize:Apl.Science:2016}.
\sloppy 
As is well-known, in the last 10 years Machine Learning (ML) and specifically Neural Networks (NN) have seen an unprecedented explosion in research, in both theoretical topics and in their application to many diverse fields of study.Applications of such techniques to modeling musically-relevant electrical circuits has seen much research in the last several years \cite{Parker2019modelling, Wright2019RNN, Damskagg2019ICASSP}. Such circuits can generally be thought to be governed by ODEs. The application of NNs to model PDE-governed systems has also seen some research, but strongly in the domain of scientific computing \cite{Li2020,Raissi2019PINN}. However, with some exceptions \cite{Gabrielli2017}, there is little work on the use of NNs to model musically relevant PDE-governed systems at audio-rates. 
\fussy
In this work we attempt to make some first steps into this domain. 
In particular, we extend on network types previously proposed in the scientific-computing domain for the modeling of PDE-governed systems \cite{Li2020} and propose novel network structures with improved capability to be applied in the audio domain. 
The proposed network structure is perfectly suited for the modeling of oscillating acoustical systems which we show by revealing the similarities to established techniques for physical modeling based on modal synthesis. 
Finally, we demonstrate the viability of the proposed structures by the modeling of three example systems. All required code for reproducing the results are provided online \cite{web}. 

In Sec.~\ref{sec:fno} we review current research on the application of NNs to solve PDEs, and introduce the proposed network structures. Moreover, we review the relevant formalism of the FTM and discuss its connection to the proposed NN structure.
In Sec.~\ref{sec:models}, we introduce three acoustical systems which are used for the evaluation of the proposed NN structures.
In Sec.~\ref{sec:evaluation} we first explain the NN training procedure applied and discuss the  performance of our proposed NN structure compared to analytical solutions of the example systems. Sec.~\ref{sec:conclusion} concludes the paper.

\section{Modeling PDEs using NN structures}
\label{sec:fno}
The modeling of ODE-governed systems via the application of machine learning is relatively trivial to construct, given that we can write such a system in the general case as follows
\begin{equation}
    \dot{\mathbf{u}}(t) = g(\mathbf{u}(t)),
\label{eq:ode}
\end{equation}
where $\mathbf{u}$ is an $n$-dimensional vector of scalar-valued states, and $g$ is an arbitrary pointwise mapping $\mathbb{R}^n \rightarrow \mathbb{R}^n$. This formulation even encompasses implicitly defined ODE systems, given that we do not need to know $g$ in analytical form, and hence it can be assumed to resolve the implicit relationship. 

From this formulation, it is clear that we can learn the dynamic of any ODE system by approximating the function $g$ with a neural network. This learned function can then be used directly within a standard ODE solver \cite{Raissi2018Multistep}. Alternatively, a \emph{recurrent neural network} (RNN)-like structure can be trained to reproduce the progression between sampled values of $\mathbf{u}$ \cite{Parker2019modelling}.

Modeling PDEs is conceptually a little more difficult, as they deal with the evolution of higher-dimensional distributed states rather than 0-dimensional scalar states, and hence cannot be characterized by a pointwise mapping. Initial work in this area, known as \emph{physically informed neural networks} (PINN), approached this problem by using a neural network to directly learn the solution manifold of the system as a function of coordinates. Such an approach has two major downsides in the context of audio-focused physical modeling. Firstly it requires the governing PDE to be already known. Secondly, it learns only the solution to a particular set of initial conditions, and therefore cannot be used flexibly to reproduce the response to arbitrary excitations.

Recent work has framed the problem differently, by formulating time-dependent PDE-systems analogously to \eqref{eq:ode} as follows
\begin{equation}
    \dot{\mathbf{u}}(\mathbf{x},t) = \mathcal{G}(\mathbf{u}(\mathbf{x},t)),
    \label{eq:fno:2}
\end{equation}
where $\mathbf{u} = \left(u_i\right)_{i = 0}^N$ is a vector of space and time-dependent states, e.g., physical quantities of the underlying system, on a bounded $n$-dimensional region $\mathbf{x}\in V \subset \mathbb{R}^n$, and $\mathcal{G}$ is an operator mapping between function spaces.

The machine-learning problem can be formulated by discretizing $\mathbf{u}$ in space and time. In particular, we apply spatial sampling on an arbitrary grid of discrete sampling points $\mathbf{x}_{\mathbf{n}}$ with $\mathbf{n}\in\mathbb{I}$, where the index set $\mathbb{I}$ is chosen such that $\mathbf{x}_\mathbf{n}\in V$, and temporal sampling at regular intervals in time, i.e., $t = kT$ with discrete-time index $k$ and sampling interval $T$. The temporal and spatial discretization of $\mathbf{u}$ in \eqref{eq:fno:2} yields the tensor
\begin{equation}
    \mathbf{U}[\xn,k] = \left(u_i[\xn,k] \right)_{i=0}^N.
    \label{eq:fno:3}
\end{equation}
The discretization of $\mathbf{u}$ allows us to define a discretized version of the PDE \eqref{eq:fno:2} in terms of the tensor \eqref{eq:fno:3} as follows
\begin{equation}
    \mathbf{U}[\xn,k+1] = \hat{\mathcal{G}}\left(\mathbf{U}[\xn,k]\right),
    \label{eq:fno:4}
\end{equation}
where $\hat{\mathcal{G}}$ is a discretized version of the original operator $\mathcal{G}$. 
Approximating the operator $\hat{\mathcal{G}}$ by a NN is known as the \emph{neural operator} approach \cite{Li2020}. In the following, we use the shorthand $\bm{U}^{k}$ to denote the tensor at time $k$, and $u^k_{ij\dots}$ denotes its elements, where $i$ is the index of the physical quantity, $j\dots$ denote the $n$ spatial variables.

There are many options for what type of network to use for this approximation. The first impulse might be to flatten the tensor $\mathbf{U}$ and use a standard dense network. This approach would make no assumptions about the relationship between elements of the tensor. This could work theoretically, but discards a number of structural priors that can be used to inform the network design and to ease training. The dimensions of the tensor $\mathbf{U}$ in \eqref{eq:fno:2} representing the discretized spatial domains are made up of progressively sampled values from continuous functions. The values along these dimensions are strongly related as they embed some concept of locality and ordering. Instead of using a fully dense transformation to operate along these dimensions, it is therefore more sensible to use a structure that contains these assumptions and transforms the data with some type of kernel. Previous work has investigated a number of approaches to this, including \emph{graph neural networks} \cite{Kovachki2021}, \emph{convolutional neural networks} with small kernels \cite{Kovachki2021}, and transforming into the spatial Fourier domain \cite{Li2020}. This last technique has seen the greatest success, and is known as the \emph{fourier neural operator} (FNO) approach.

These kernel-type methods of approximating the operator can be considered to be closely related to general kernel methods for the solution of PDEs such as Green's Functions or the eigenfunctions of the spatial differentiation operator concealed in~\eqref{eq:fno:2}~\cite{Trautmann2003DigitalMethod,Schafer2019SimulationModels}.
These relations are discussed in further detail in Sec.~\ref{sec:refFNO-Kern}.

FNO-based techniques are significantly better suited than the PINN approach for audio-focused physical modeling, but have so far only been applied in the context of scientific computing. Compared to many typical scientific computing problems, modeling audio-range behaviour requires a system to reproduce dynamics over a wide frequency range, and across long time-scales relative to the temporal sampling frequency. Initial experiments showed that taking the existing FNO-based structures and applying them directly to audio-focused problems did not produce ideal results - in particular the modeled behaviour would degenerate after a relatively small number of timesteps. The goal of this work was therefore to improve the existing FNO-based structures in the aspects important to audio-range modeling.

\subsection{Fast convolution layer}
Starting from the n-dimensional spectral convolution layer proposed in the context of the FNO \cite{Li2020, Kovachki2021}, we make some modifications for more generalized usage. Firstly, we discard the concept of zeroing some of the bins of the FFT of the data. This zeroing is effectively just a crude lowpass filter implemented in the Fourier domain, and is not beneficial in the general use-case. We also add correct padding of the spatial dimensions to ensure that the layer is performing non-cyclic convolution. 

\sloppy
The operation of the layer on the internal state tensor $\bm{H}$ with elements $h_{\nu j\hdots}$ can be written as follows
\begin{equation}
    \mathcal{S}(h_{\nu j\hdots}) = \mathcal{\tilde{F}}^{-1}\left[\sum_\kappa A_{\nu \kappa j\hdots}\mathcal{\tilde{F}}(h_{\kappa j\hdots})+b_{\nu j \hdots}\right],
    \label{eq:fftlayer}
\end{equation}
where $\mathcal{\tilde{F}}$ denotes an operator encompassing padding, applying the FFT and truncating the spectrum to remove negative frequency components. Conversely $\mathcal{\tilde{F}}^{-1}$ encompasses reconstructing the negative freq-uency components from the transformed positive freq-uency components, applying the inverse FFT, and then truncating to remove padding.  Note that this linear transformation with $\bm{A}$ and $\bm{b}$ is fully dense with respect to input and output channels but diagonal in terms of frequency bins. The elements $A_{\nu \kappa j\hdots}$ and $b_{\nu j \hdots}$ of this transformation are complex, and are the trainable parameters of the layer. 
With these modifications, the layer can be thought of as a standard convolutional layer with kernels fixed to the width of the domain, but implemented using the well-established method of fast-convolution via multiplication in the Fourier domain. This has an advantage in computational complexity, and also re-contextualizes the training problem by moving parameters from the spatial domain into the Fourier domain.
\fussy
It should be noted that whilst the general discretization shown in \eqref{eq:fno:3} allows arbitrary spatial sampling grids, the use of the FFT to implement convolution could be argued to enforce a regular rectilinear spatial grid. This is certainly true in the linear case, but in the context of stacked layers with non-linearites, this may not be true. Nonetheless, it is reasonable to assume that compensating for a non-regular grid may use significant capacity of the network. A proper investigation of this topic is left to future work, with only data sampled on rectilinear spatial grids considered here.

\subsection{FNO-derived RNN structures}
\sloppy
Previously presented FNO structures have dealt with time-evolu\-tion by learning mappings from spatially sampled states at multiple input time-steps \cite{Li2020}, or single input time-steps \cite{markov} to the next time-step. These structures can be thought of as a type of teacher-forced RNN, but were not previously discussed as such. In this work we introduce several related structures that are designed explicitly as RNNs, and trained using standard RNN training techniques such as \emph{back propagation through time} (BPTT).
\fussy


\subsubsection{Fourier Recurrent Neural Network (FRNN)}
\begin{figure}
    \centering
    \includegraphics[width = 0.9 \columnwidth]{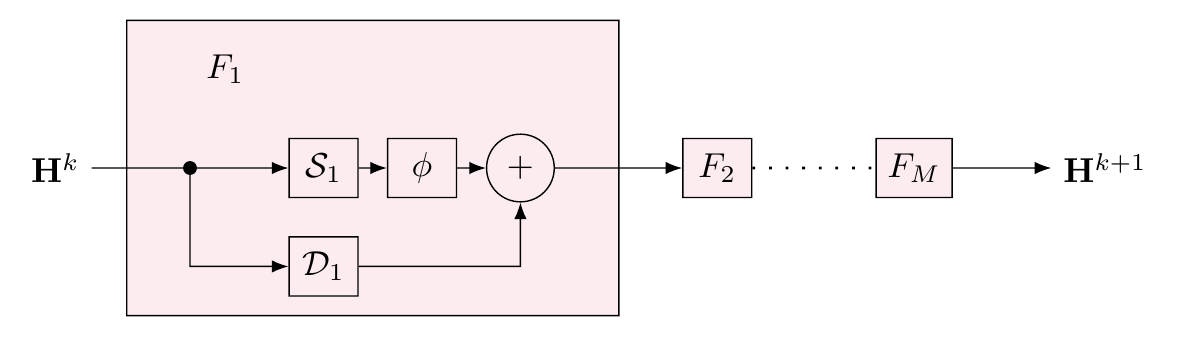}
    \caption{Block diagram showing the recurrent cell of the FRNN structure  (dependencies of $\mathbf{H}$ on discrete time index $k$ are denoted in the superscript for brevity).}
    \label{fig:frnn}
\end{figure}
This structure is shown in Fig.~\ref{fig:frnn}. It consists of a recurrent structure that maps sequentially between time-steps of the spatially-sampled internal states of the network, through a variable number of mapping blocks
\begin{equation}
\mathbf{H}^{k+1} = F_M \circ F_{M-1} \circ \hdots F_1(\mathbf{H}^k),
\label{eq:nn_multilayer}
\end{equation}
where $\mathbf{H}$ is a tensor containing the internal states of the system at all spatial-sampling points, and the mapping block is given by
\begin{equation}
    F_m(\mathbf{H}) = \phi(\mathcal{S}_m(\mathbf{H})) + \mathcal{D}_m(\mathbf{H}), 
    \label{eq:nn_onelayer}
\end{equation}
where $\mathcal{S}$ is the described fast convolution layer \eqref{eq:fftlayer}, $\phi$ is an element-wise activation function such as $\text{tanh}$ or ReLU and $\mathcal{D}$ represents a weighted skip connection.

This structure is related to that presented in previous FNO literature \cite{markov} but with the skip connection positioned to prevent vanishing gradients when using BPTT on longer sequences, and without the restriction that internal states must correspond with physical states. Compared to this structure, we also do not condition the input with the coordinates of the spatial grid. This structure can also be considered to be closely related to a generic RNN, and to the STN structure proposed for modeling ODE-governed systems \cite{Parker2019modelling}.

\subsubsection{Fourier Gated Recurrent Unit (FGRU)}
\begin{figure}
    \centering
    \includegraphics[width = 0.9 \columnwidth]{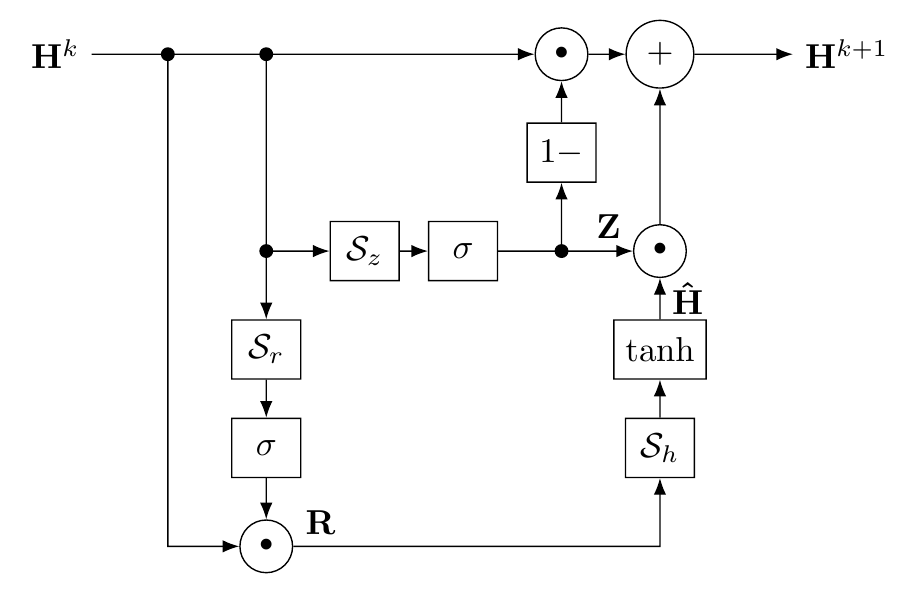}
    \caption{Block diagram showing the recurrent cell of the FGRU structure (dependencies of $\mathbf{H}$ on discrete time index $n$ are denoted in the superscript for brevity).}
    \label{fig:fgru}
\end{figure}

The \emph{gated recurrent unit} (GRU) \cite{GRU} is an RNN structure that has seen great success in a variety of tasks from language modeling to black-box modeling of electrical circuits \cite{Wright2019RNN}. We propose a generalization of this structure by replacing dense layers with fast convolution layers. The structure is shown in Fig.~\ref{fig:fgru}, and can be written as follows
\begin{align}
\mathbf{Z} &= \sigma(\mathcal{S}_z(\mathbf{H}^k)), \\
\mathbf{R} &= \sigma(\mathcal{S}_r(\mathbf{H}^k)), \\
\mathbf{\hat{H}} &= \text{tanh}(\mathcal{S}_h(\mathbf{R} \odot \mathbf{H}^k)), \\
\mathbf{H}^{k+1} &= (1-\mathbf{Z}) \odot \mathbf{H}^k + \mathbf{Z} \odot \mathbf{\hat{H}},
\end{align}
where $\mathbf{H}$ is a tensor as in \eqref{eq:nn_multilayer}, \eqref{eq:nn_onelayer}, $\sigma$ is the element-wise sigmoid function, and $\mathcal{S}_z$, $\mathcal{S}_r$, $\mathcal{S}_h$ are fast convolution layers.

This structure inherits the well-known beneficial qualities of the GRU, including a linear path to avoid vanishing gradients during BPTT, the ability to effectively multiply inputs, and safety against explosive instability due to the bounded nature of the states.

\subsubsection{Training procedure}

We formulate the training problem as taking a tensor of the initial conditions of the states of the physical system, sampled on an n-dimensional regular grid, and producing a tensor of one dimension higher, representing the evolution of these spatially sampled states over a number of time-steps. This is a form of BPTT, in contrast to the single time-step to single time-step mappings previously described \cite{markov}, and hence has the advantage that the network can learn to deal with its own error as propagated through the recursion. 

\begin{figure}
    \centering
    \includegraphics[width =\columnwidth]{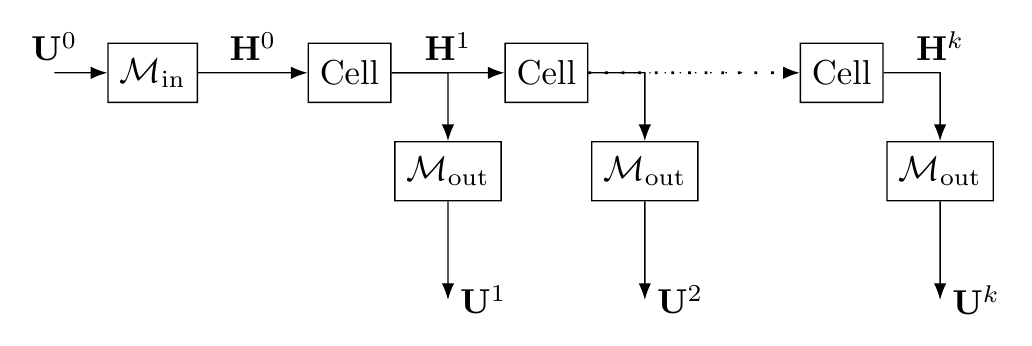}
    \caption{Block diagram showing the repeated application of an RNN cell from Fig.~\ref{fig:frnn} along with the action of the state-matching layers.}
    \label{fig:recursion}
\end{figure}

For the presented RNN structures, the number of internal states can be freely specified. This is especially important in the case of the FGRU, as it is the only mechanism by which to scale the capacity of the structure. Restricting the internal states $\bm{H}$ to correspond exactly to the physical states $\bm{U}$ of the system we are modeling would therefore be overly restrictive. Instead, we apply a process called \emph{soft state-matching}\cite{peussa2021}. This consists of defining two trainable linear maps $\bm{A}$, $\bm{b}$ and $\tilde{\bm{A}}$, $\tilde{\bm{b}}$, respectively, 
\begin{align}
        h_{\nu j\hdots}^0 &= \mathcal{M}_\mathrm{in}(u_{i j\hdots}^0) = \sum_\kappa A_{\nu\kappa}u_{\kappa j\hdots}^0+b_\nu, \\
        u_{ij\hdots}^k &= \mathcal{M}_\mathrm{out}(h_{\nu j\hdots}^k) =\sum_\kappa \tilde{A}_{i\kappa}h_{\kappa j\hdots}^k+\tilde{b}_{i},
        \label{eq:nn_inoutput}
\end{align}
where $h_{\nu j\hdots}$ are the elements of the network's internal state tensor $\mathbf{H}$ and $u_{ij\hdots}$ are the elements of the tensor $\mathbf{U}$ containing the states of the system being modeled. The superscript $k$ denotes time-step. These maps translate from the dimensionality of the states of the system being modeled to that of the network, and vice-versa. The first mapping is applied to translate the initial conditions to the initial internal states of the network, and the second mapping is used to translate the states of the network to the states of original system. With the addition of these mappings, the network can be thought of as operating in a higher-dimensional latent space which contains a linear embedding of the original state-space. The combination of recursive training via BPTT with these mapping layers is shown in Fig.~\ref{fig:recursion}.


\subsection{Relation between FNO structures and Kernel Methods}
\label{sec:refFNO-Kern}

As previously mentioned, there is a strong relation between the FNO approach and general kernel methods for the solution of PDEs in the form of \eqref{eq:fno:2}. In the following we investigate the FTM, as a kernel method for finding analytical solutions to \eqref{eq:fno:2}, the relation to FNOs and to the proposed FNO-derived RNN Structures. 

\subsubsection{Functional Transformation Method (FTM)}
\label{sec:ftm}
The FTM is a powerful method for modeling the oscillation of acoustic systems in terms of transfer functions and a detailed description of its derivation and application can be found, e.g., in \cite{Trautmann2003DigitalMethod, Schafer2019SimulationModels,petrausch:SMMSP:2004}. 
For example, the method has been applied for the modeling of strings (see, e.g., \cite{schaefer:dafx:2016,Trautmann2003DigitalMethod}), membranes (see, e.g., \cite{rabenstein:ieee:2010,Han2020}, and room acoustics (see, e.g., \cite{Petrausch2005SimulationMethod,Schafer2020ASynthesis}).
The basic idea of the FTM is to transform the mathematical description of an acoustical system in terms of PDEs, e.g., \eqref{eq:fno:2}, into a discrete-time simulation model by the application of functional transformations for time and space dependencies. 
Consider a vector valued PDE, which is a specific realization of the general PDE in \eqref{eq:fno:2}
\begin{align}
    \mathbf{C}\dot{\mathbf{u}}(\mathbf{x},t) &= \mathbf{L}\mathbf{u}(\mathbf{x},t), &\mathbf{x}\in V, \, t > 0,\label{eq:ftm:1}
\end{align}
defined on spatial domain $V$ and its boundary $\partial V$. In PDE \eqref{eq:ftm:1}, matrix $\mathbf{C}\in\mathbb{R}^{N\times N}$ is a capacitance matrix and $\mathbf{L}$ is a $N\times N$ sized spatial differentiation operator. 
The state vector $\mathbf{u}\in\mathbb{R}^{N\times 1}$ contains $N$ physical quantities of the underlying acoustical system such as deflection, sound pressure or particle velocities (see, e.g., \cite[Eqs.~(26), (34)]{Schafer2020ASynthesis}). 
To complete PDE \eqref{eq:ftm:1}, we assume a set of homogeneous boundary conditions defined on $\partial V$, cf. \cite{petrausch:SMMSP:2004}, and an initial condition for the vector $\mathbf{u}$ at $t = 0$, i.e., $\mathbf{u}(\mathbf{x},0) = \mathbf{u}_\mathrm{i}(\mathbf{x})$.

In order to derive an FTM model for \eqref{eq:ftm:1}, $\mathbf{u}$ is expanded into an infinite set of bi-orthogonal basis functions $\Kprim_\mu\in\mathbb{C}^{N\times 1}$ and $\Kadj_\mu\in\mathbb{C}^{N\times 1}$ for the operator $\mathbf{L}$, where the dedicated eigenvalues $s_\mu$, $\mu\in\mathbb{N}_0$ constitute the discrete spectrum of $\mathbf{L}$ \cite{churchill:1972}. With basis function $\Kprim_\mu$ and $\Kadj_\mu$ a forward and inverse Sturm-Liouville transform (SLT) are defined as follows 
\begin{align}
	\bar{u}_\mu(t) &= \mathcal{T}\{\mathbf{u}(\mathbf{x},t)\} = \int_{V}\Kadj_\mu\her(\mathbf{x})\mathbf{C}\mathbf{u}(\mathbf{x},t)\dint{\mathbf{x}},\label{eq:ftm:4}\\
	\mathbf{u}(\mathbf{x},t) &= \mathcal{T}^{-1}\{\bar{u}_\mu(t)\}=\sum_{\mu = 0}^{\infty} \frac{1}{N_\mu}\bar{u}_\mu(t) \Kprim_\mu(\mathbf{x}). 
	\label{eq:ftm:5}
\end{align}
Forward SLT \eqref{eq:ftm:4} expands $\mathbf{u}$ into the basis functions $\Kadj_\mu$ yielding the expansion coefficients $\bar{u}_\mu$. 
In the context of acoustics, \eqref{eq:ftm:4} expands a system into its modes and the time-dependent expansion coefficients $\bar{u}_\mu$ describe their temporal evolution. 
The inverse SLT in \eqref{eq:ftm:5} represents $\mathbf{u}$ as series expansion with basis functions $\Kprim_\mu$ and scaling factors $N_\mu = \int_V \Kadj_\mu\her\mathbf{C}\Kprim_\mu\dint{\mathbf{x}}$. 
Further properties of \eqref{eq:ftm:4} and \eqref{eq:ftm:5} such as the existence of a differentiation theorem and the bi-orthogonality of $\Kprim$ and $\Kadj$ are discussed in detail in \cite[Sec.~4.7.3]{Schafer2019SimulationModels}.
We note that the shape of $\Kprim$ and $\Kadj$ depends on the spatial shape of the underlying system, e.g., for geometrically simple systems basis functions are often trigonometric functions, Bessel functions or combinations thereof \cite[Sec.~4.7.4]{Schafer2019SimulationModels}. 

Application of the forward SLT \eqref{eq:ftm:4} to PDE \eqref{eq:ftm:1} and exploiting the differentiation theorem \cite[Eq.~(18)]{Schafer2020ASynthesis} leads to expansion coefficients defined in terms of Laplace transfer functions 
\begin{align}
    &\bar{u}_\mu(t) = \mathrm{e}^{s_\mu t}\bar{u}_{\mathrm{i},\mu} &\laplace && \bar{U}_\mu(s) = \frac{1}{s-s_\mu}\bar{u}_{\mathrm{i},\mu}, 
    \label{eq:ftm:6}
\end{align}
where the coefficients $\bar{u}_{\mathrm{i},\mu}$ follow from the expansion of the initial values $\mathbf{u}_\mathrm{i}$, i.e.,  $\bar{u}_{\mathrm{i},\mu} = \mathcal{T}\{\mathbf{u}_\mathrm{i}(\mathbf{x})\}$.

Truncating the number of eigenvalues $s_\mu$ to be finite, i.e., $\mu = 0, \dots, Q-1$, and transforming \eqref{eq:ftm:5} and \eqref{eq:ftm:6} into the discrete-time domain allows the formulation of the FTM model in terms of a vector valued state space description (SSD) \cite{Schafer2019SimulationModels,Schafer2020ASynthesis}
\begin{align}
    \bar{\mathbf{u}}[k+1] &= \mathrm{e}^{\As T}\bar{\mathbf{u}}[k] +T\bar{\mathbf{u}}_\mathrm{i}\delta[k], \label{eq:ftm:7}\\
    \mathbf{u}[\mathbf{x},k] &= \Cs(\mathbf{x})\bar{\mathbf{u}}[k], \label{eq:ftm:8}
\end{align}
with the discrete time index $k$, sampling interval $T$, i.e., $t=kT$, and the discrete-time Dirac delta function $\delta[k]$. \textit{State equation} \eqref{eq:ftm:7} is a vector valued discrete-time version of \eqref{eq:ftm:6}, where the vector $\bar{\mathbf{u}} = \left(\bar{u}_\mu \right)_{\mu = 0}^{Q-1}$ contains the expansion coefficients, the diagonal matrix $\As\in\mathbb{C}^{Q\times Q}$ contains the eigenvalues $s_\mu$ on its main diagonal and $\bar{\mathbf{u}}_\mathrm{i} = \left(\bar{u}_{\mathrm{i},\mu} \right)_{\mu = 0}^{Q-1}$. \textit{Output equation} \eqref{eq:ftm:8} is a clever reformulation of the inverse SLT in \eqref{eq:ftm:5} by representing the truncated sum by a matrix-vector multiplication with the transformation matrix 
\begin{align}
    \Cs(\mathbf{x}) = \left[\frac{1}{N_0}\Kprim_0(\mathbf{x}), \,\dots, \frac{1}{N_{Q-1}}\Kprim_{Q-1}(\mathbf{x}) \right]. \label{eq:ftm:9}
\end{align}

\subsubsection{Relation between FTM and FNO-derived RNN Structures}
\label{subsec:fnoftm}

The FTM formulation allows some interpretation of the fast convolution layer proposed in the FNO-derived RNN structure. The state equation \eqref{eq:ftm:7} describes the time-evolution of the modes phase and amplitude $\bar{\mathbf{u}}$. The state transition matrix $\mathrm{e}^{\As T}$ is diagonal, such that time-evolution of each mode state $\bar{\mathbf{u}}$ is decoupled. 

For a set of discrete spatial sampling points $\xn$ with $n \in \mathbb{I}$ and discrete time $k$, the transformation \eqref{eq:ftm:4} and \eqref{eq:ftm:5} are 
\begin{align}
    \bar{\mathbf{u}} [k] &= \sum_{\mathbf{n}\in\mathbb{I}}
    \Csa\her(\mathbf{x}_\mathbf{n}) \mathbf{C} V_\mathbf{n}\, \mathbf{u}[\mathbf{x}_\mathbf{n},k],
    \label{eq:dis:5}\\
    \mathbf{u}[\mathbf{x}_\mathbf{n},k] &=
    \Cs(\mathbf{x}_\mathbf{n})\,\bar{\mathbf{u}} [k],
    \label{eq:dis:6}
\end{align}
where $V_\mathbf{n}$ denotes the finite volume element replacing the infinitesimal volume element $\dint{\mathbf{x}}$ and $\Csa$ is defined by the eigenfunctions $\Kadj$ similar to \eqref{eq:ftm:9}. For geometrically simple examples, such as strings and rectangles, the underlying basis functions $\Kprim_\mu$ are trigonometric. In a linear time-invariant system, the internal modes can be recovered readily from the time evolution of the spatial variables. A relation between two consecutive time steps in the space domain follows by inserting~\eqref{eq:ftm:7} and~\eqref{eq:dis:5} for $k>0$ into~\eqref{eq:ftm:8}
\begin{align}
\mathbf{u}[\mathbf{x}_\mathbf{n},k+1] &=\sum_{\mathbf{\nu}\in\mathbb{I}} \mathbf{G}(\mathbf{x}_\mathbf{n}|\mathbf{x}_\mathbf{\nu}) \mathbf{u}[\mathbf{x}_\mathbf{\nu},k]
    \label{eq:poles:1}
\end{align}
with the matrix in the form of an eigenvalue decomposition
\begin{align}
\mathbf{G}(\mathbf{x}_\mathbf{n}|\mathbf{x}_\mathbf{\nu}) 
= 
\Cs(\mathbf{x}_\mathbf{n}) \mathrm{e}^{\As T} \Csa\her(\mathbf{x}_\mathbf{\nu}) \mathbf{C} V_\mathbf{n}.
    \label{eq:poles:2}
\end{align}
Note that \eqref{eq:poles:1} is the linear version of the general discretized PDE \eqref{eq:fno:4} and $\mathbf{G}(\mathbf{x}_\mathbf{n}|\mathbf{x}_\mathbf{\nu})$ resembles a discrete version of the Green's function of \eqref{eq:ftm:1}.
Transition matrix $\mathbf{G}$ can be numerically recovered from the spatial variables by solving a least squares problem in \eqref{eq:poles:1} on a sufficiently long time frame. The eigenvalue decomposition of $\mathbf{G}$ yields then eigenvalues $\mathrm{e}^{\As T}$.

Similar operations are performed by the fast convolution layer in \eqref{eq:fftlayer}. An FFT transforms the spatial variables into a transform domain with complex exponential basis functions, thus also trigonometric functions such as in $\Cs$. The processing step in \eqref{eq:fftlayer} applies a diagonal transition matrix $A_{\nu \kappa j\hdots}$ to the transform-domain variables. The result is transformed back to the spatial variables via the inverse FFT. There are also notable differences between the given approaches. The fast convolution layer applies a bias term, i.e., $b_{\nu j \hdots}$ in \eqref{eq:fftlayer}. The FNO-derived RNN structure typically concatenates multiple layers (see \eqref{eq:nn_multilayer}), uses non-linear activation functions and skip connections in-between (see \eqref{eq:nn_onelayer}). Because of the mapping performed by the first and last network layers as in \eqref{eq:nn_inoutput}, the NN recombines and expands the spatial variable inputs and outputs.


\section{Acoustic Systems}
\label{sec:models}
In this section we introduce a number of acoustic systems which are used to test the viability of the proposed NN structures.
First, we investigate two linear systems, -- a \textit{lossy dispersive string} and a \textit{2D wave equation}. 
For both systems we employ well investigated models obtained by the FTM as introduced in Sec.~\ref{sec:ftm}. 
Second, we investigate a \textit{tension modulated string} as a non-linear system which we solve, after a few modifications, by a non-linear ODE-solver in Python. 

The models presented have been well investigated previously therefore we just present the parts necessary for a comprehensible presentation. We refer the reader to the relevant literature where appropriate.

\subsection{Linear Lossy Dispersive String}
\label{subsec:model:string}
The oscillation of a vibrating string of length $\ell$, i.e., $0 \leq x \leq \ell$ at times $t > 0$ can be described by a PDE in terms of the string deflection $u_0(x,t)$ as follows \cite{Bensa:JASA:2003} 
\begin{align}
\rho_\mathrm{s} A u_1(x,t) &+ EI u_0''''(x,t) - T_{\mathrm{s}0} u_0''(x,t) \nonumber\\ 
&+ d_1 u_1(x,t) - d_3 u_1''(x,t) = 0, \label{eq:string:1}
\end{align}
where partial derivatives for space are denoted by a prime, i.e., $\frac{\partial u}{\partial x} = u'$, and velocity by $u_1$.
The oscillation of the string in \eqref{eq:string:1} is influenced by several physical parameters. In particular, $\rho_\mathrm{s}$ denotes the string density, $A$ is the cross section area, $I$ is the moment of inertia and $T_{\mathrm{s}0}$ is the constant tension of the string. 
Constant $E$ denotes Young's modulus. The parameters $d_1$ and $d_3$ introduce frequency-independent and frequency-dependent damping into the oscillation of the string. 
As boundary conditions we assume that the string is simply supported, which requires that the deflection $u_0$ and its second derivative $u_0''$ are zero at both ends, see \cite{schaefer:dafx:2016}. At $t = 0$, deflection is defined by an initial value \mbox{$u_0(x,0) = u_{\mathrm{i}}(x)$}.

The derivation of an FTM model for the lossy dispersive string has been discussed, e.g., in \cite{petrausch:SMMSP:2004,schaefer:dafx:2016,Schafer2020ASynthesis, Trautmann2003DigitalMethod} for different applications in sound synthesis. To this end, we don't show the exact formulas for all components of the FTM model, instead we refer the reader to the most recent FTM references:
\begin{itemize}
\setlength{\itemsep}{0pt}
\setlength{\parskip}{0pt}
    \item To derive the state transition matrix $\mathrm{e}^{\As T}$ in \eqref{eq:ftm:7}, we used the eigenvalues $s_\mu$ from \cite[Eq.~(37)]{Schafer2020ASynthesis}.
    \item The eigenfunctions $\Kprim_\mu$ in $\Cs$ in \eqref{eq:ftm:9} and $\Kadj_\mu$ for the expansion in \eqref{eq:ftm:4} can be found in \cite[Eq.~(39)]{Schafer2020ASynthesis}.
\end{itemize}
The vector of states $\mathbf{u}$, c.f. \eqref{eq:ftm:1}, \eqref{eq:fno:2}, comprises string deflection $u_0$ and velocity $u_1$, i.e., $\mathbf{u}=\left[u_0,\,u_1\right]\tran$.
For training and evaluation we use different initial values $u_\mathrm{i}$, i.e., a delta impulse, a smooth excitation by a raised cosine, and a random initial condition.

\subsection{2D Wave Equation}
\label{subsec:model:room}

The evolution of sound pressure $u_0$ and particle velocities $\mathbf{u}_\mathrm{v} = \left[u_1,\,u_2\right]\tran$ in a bounded 2-dimensional (2d) region of size $0\leq x \leq L_x$, $0\leq y \leq L_y$ is described by the 2d wave equation 
\begin{align}
    \rho_0 \Dot{\mathbf{u}}_\mathrm{v}(x,y,t) + \mbox{grad}\,u_0(x,y,t) &= 0, \label{eq:room:1}\\
    \rho_0\, c_0^2\, \mbox{div}\,\mathbf{u}_\mathrm{v}(x,y,t) + \Dot{u}_0(x,y,t) &= 0,\label{eq:room:2}
\end{align}
where $\rho_0$ denotes the density of air and $c_0$ is the speed of sound. The operators $\mbox{grad}$ and $\mbox{div}$ denote gradient and divergence in 2D Cartesian coordinates, respectively. For the boundary conditions we assume fully reflective boundaries, i.e., particle velocities $\mathbf{u}_\mathrm{v}$ vanish at the boundaries. At $t = 0$, sound pressure $u_0$ is defined by an initial value, i.e., $u_0(x,y,0) = u_\mathrm{i}(x,y)$. The 2D wave equation \eqref{eq:room:1}, \eqref{eq:room:2} and its extension to 3D are frequently used in room acoustics~\cite{Blauert2008AcousticsEngineers,Kuttruff2016RoomAcoustics}. 
The components required to establish a FTM model for the 2D wave equation are as follows: 
\begin{itemize}
\setlength{\itemsep}{0pt}
\setlength{\parskip}{0pt}
    \item To derive the state transition matrix $\mathrm{e}^{\As T}$ in \eqref{eq:ftm:7}, we used the eigenvalues $s_\mu$ from \cite[Eq.~(30)]{Schafer2020ASynthesis}.
    \item The eigenfunctions $\Kprim_\mu$ in $\Cs$ in \eqref{eq:ftm:9} and $\Kadj_\mu$ for the expansion in \eqref{eq:ftm:4} can be found in \cite[Eq.~(31)]{Schafer2020ASynthesis}.
\end{itemize}
The vector of states $\mathbf{u}$, c.f. \eqref{eq:ftm:1}, \eqref{eq:fno:2}, comprises sound pressure $u_0$ and particle velocities $\mathbf{u}_\mathrm{v}$, i.e., $\mathbf{u}=\left[u_0,\,u_1,\,u_2\right]\tran$.
Similar to the string we use different types of initial conditions $u_\mathrm{i}$, i.e., a 2d delta impulse, and a random sound pressure distribution. 

\subsection{Non-linear Tension Modulated String}
\label{subsec:model:tstring}
As a non-linear acoustical system we consider a tension modulated string \cite{Trautmann2000SoundTransformations,Trautmann2003DigitalMethod,Bilbao:dafx:2004,Avanzini:JASA:2012, Tolonen2000ModelingStrings}. Tension modulation is a non-linear phenomenon which affects the pitch of the string. In particular, when the string is deflected from the rest position, the arc length measured along the string is larger than the string length $\ell$ at rest. This additional ``stretching'' leads to an increased tension and subsequently to an increased pitch. 

The PDE describing the oscillation on a tension modulated string is similar to the linear case \eqref{eq:string:1}, except for the tension which depends on $u_0(x,t)$ and is defined as follows
\begin{align}
    T_\mathrm{s}(u_0) = T_{\mathrm{s}0} + T_{\mathrm{s}1}(u_0),
    \label{eq:stringten:1}
\end{align}
where $T_{\mathrm{s}0}$ is the tension of the string at rest as defined in \eqref{eq:string:1}. The additional tension $T_{\mathrm{s}1}(u_0)$ arising from the additional arc length $\Delta l_\mathrm{str}(u_0)$ of the string is derived from Hooke's law  \cite{Tolonen2000ModelingStrings} 
\begin{align}
     T_{\mathrm{s}1}(u_0) = EA\frac{\Delta l_\mathrm{str}(u_0)}{\ell} = \frac{EA}{2\ell}\int_{0}^\ell\left(u_0'(x,t)\right)^2\mathrm{d}x.
     \label{eq:stringten:2}
\end{align}
In contrast to the linear example systems, we derived a numerical model for the tension modulated string. In particular, we employed a two-staged procedure to obtain numerical solutions for the deflection $u_0$ and velocity $u_1$ of the tension modulated string:
\begin{itemize}
\setlength{\itemsep}{0pt}
\setlength{\parskip}{0pt}
    \item[1)] First we decompose PDE \eqref{eq:string:1}, extendend by the non-linear tension \eqref{eq:stringten:1}, into a set on non-linear ODEs by the application of a Fourier-Sine transformation~\cite{Bilbao:dafx:2004}. 
    \item[2)] The system of ODEs is converted into a vector formulation of size $M$ ODEs, solved numerically in Python. 
\end{itemize}
Similar to the linear string we employ different types of initial conditions for $u_0$, i.e., a delta impulse and a random initial condition, and the same state vector $\mathbf{u}$.

\section{Evaluation}
\label{sec:evaluation}
\fussy

\begin{figure}
    \centering
        \includegraphics[width = 0.9 \columnwidth]{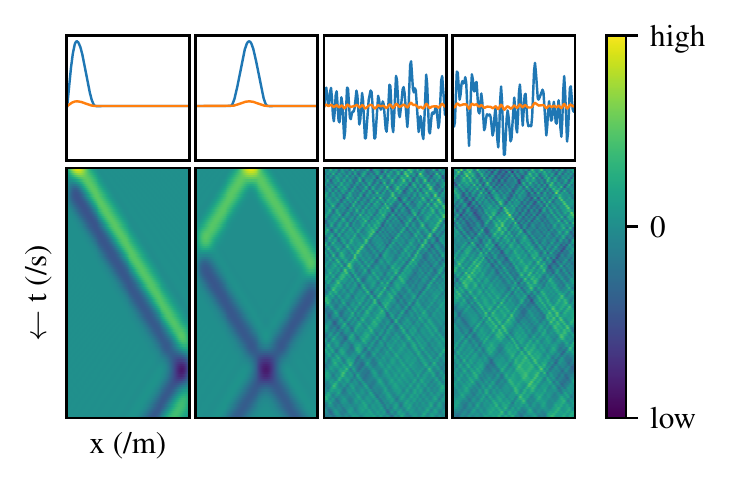}
    \caption{Examples of training pairs in the case of the linear string model. The top row shows the initial conditions, in this case of two states. The bottom row shows the evolution of the first state over a small time period. The dataset includes both states.}

    \label{fig:1d_string_examples}
\end{figure}

The networks described in Sec.~\ref{sec:fno} were implemented in the PyTorch framework\cite{pytorch}, and are available at the accompanying website\cite{web}. A reference implementation of the previous Markov-FNO approach was also adapted from existing code, with the addition of support for BPTT. These networks were then trained on datasets generated from the models described in \ref{sec:models}. These datasets consist of 1024 pairs of a set of initial conditions and the response over a set period of time. The initial conditions are split between two categories. Half consist of the response to randomly positioned impulses or plucks, and the other half the response to setting the initial conditions to random values across the domain. The datasets are normalized to have unit variance. One tenth of each dataset is retained for validation. Further physical parameters of the datasets are given at the accompanying website \cite{web}. Fig~\ref{fig:1d_string_examples} shows an example of the types of training pairs provided in the case of the 1d string.


\subsection{Training methodology}
Training was conducted using the AdamW optimizer \cite{AdamW} with default parameters, and the 1-cycle learning-rate scheduling scheme \cite{superconvergence} modulating from a learning rate of $10^{-4}$ to $10^{-3}$, with MSE between the target and predicted output sequences as the objective. Training was conducted for 5000 epochs with batch size set individually for each dataset in order to maximize GPU memory usage. The training was conducted on cloud-hosted virtual machines equipped with NVIDIA T4 GPUs. All training code is made available at the accompanying website, which also documents the used hyper-parameters \cite{web}. Network capacities were set to be approximately equal between the different network architectures. In practice, this means using the same number of internal states, with 3 stacked layers in the FRNN and reference FNO model used to match the non-variable 3 layers in the FGRU.

\subsection{Results}
In the below table we give MSE values for the proposed models and the reference model, validated on a reserved subset of the data not seen during training: 
\begin{center}
\begin{tabular}{c|c c c c}
  & FGRU & FRNN & Ref. \\ \hline
 1d linear string & \textbf{7.79e-3} & 1.43e-2 & 3.50e-1 \\ 
 1d nonlinear string &\textbf{2.27e-3} & 2.63e-2 & 1.06 \\
 2d wave equation & 1.86e-2 & 1.54e-2 & \textbf{1.12e-4}
\\
\end{tabular}
\label{tab:validation}
\end{center}

\subsubsection{Linear Lossy Dispersive String}
\begin{figure}
    \centering
        \includegraphics[width = 0.9 \columnwidth]{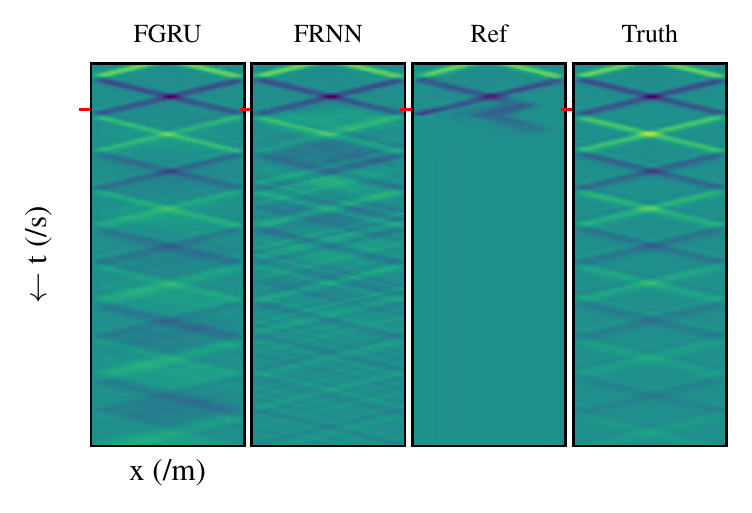}
    \caption{Deflection of the linear string over time and space as obtained by the proposed FGRU and FRNN structures, the reference FNO model and the analytical solution obtained by FTM. The color scaling is the same as given in Fig.~\ref{fig:1d_string_examples}. The red tick mark represents the duration of examples used during training.}
    \label{fig:1d_string_results}
\end{figure}

Fig.~\ref{fig:1d_string_results} shows the result of exciting the models trained on the linear string data with an impulse halfway along its length. The time-span shown is approximately 10x that seen by the models during training. As can be clearly seen, the best performing model is the FGRU, which manages to sustain accurate behavior over the majority of the time-period considered. The FRNN also decays at approximately the correct rate, but exhibits more and more dispersion after the time-span seen during training. The reference FNO model seems to fit the behavior well during the time-span seen during training, but breaks down completely after that. This ordering of accuracy is also confirmed by the validation MSE values.

\begin{figure}[h]
    \centering
        \includegraphics[width = 0.9 \columnwidth]{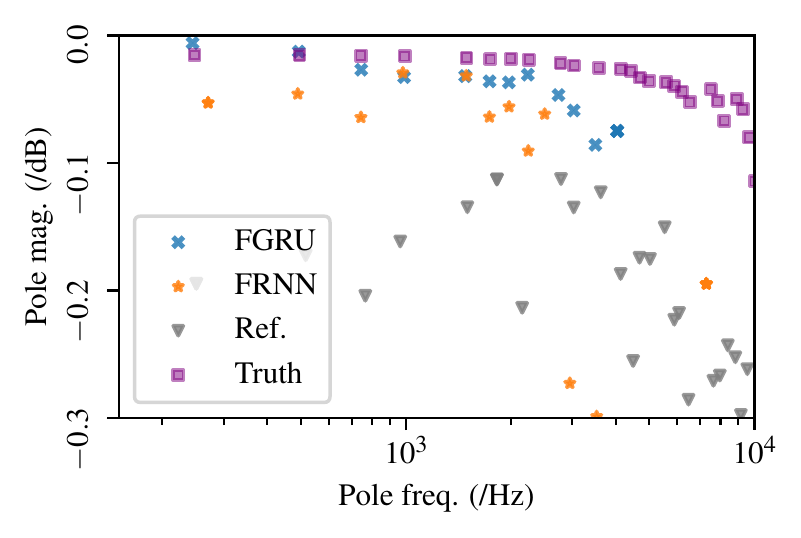}
    \caption{Estimated poles of the linear string inferred from the model outputs presented in Fig.~\ref{fig:1d_string_results}.}
    \label{fig:string_poles}
\end{figure}

We can further examine the behavior of the models by using the approximate eigenvalue decomposition derived in \ref{subsec:fnoftm}. The results are shown in Fig.~\ref{fig:string_poles}. We observe that for many of the prominent low-frequency poles of the original FTM model, the NNs recover a pole with similar frequency and magnitude. As expected from the comparison in Fig.~\ref{fig:1d_string_results}, the FGRU best matches the pole magnitudes, although the FRNN also matches them well at low frequencies. The reference FNO model seems to have estimated some pole frequencies well, but significantly damped. All of the models seems to struggle above 4kHz.

\subsubsection{Non-linear Tension Modulated String}

\begin{figure}[h]
    \centering
        \includegraphics[width = 0.9 \columnwidth]{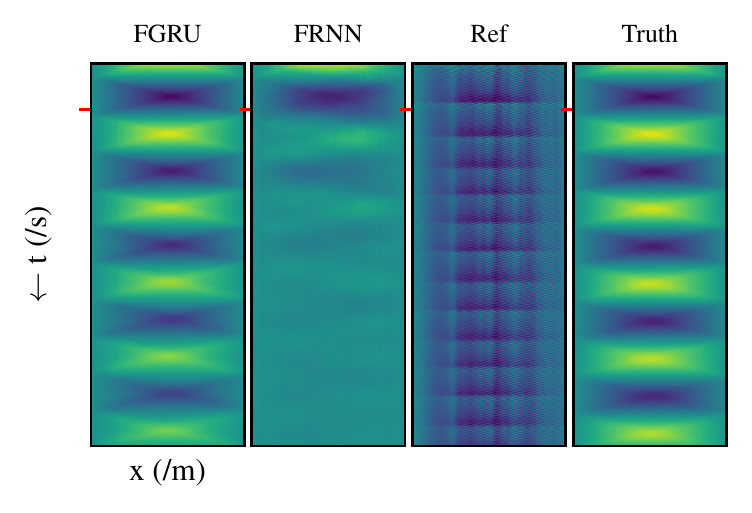}
    \caption{Deflection of the tension modulated string over time and space as obtained by the proposed FGRU and FRNN structures, the reference FNO model and the numerical solution described in Sec.~\ref{subsec:model:tstring}. The color scaling is the same as given in Fig.~\ref{fig:1d_string_examples}. The red tick mark represents the duration of examples used during training.}
    \label{fig:1d_nonlinear_string_results}
\end{figure}

Fig.~\ref{fig:1d_nonlinear_string_results} shows the result of exciting the models of the nonlinear tension modulated string with a pluck halfway along the string's length, with an initial pluck amplitude of 1mm. Again the time period shown is roughly 10x that used for training. Broadly the same hierarchy is seen as in the case of the linear string in Fig.~\ref{fig:1d_string_results}. The result from FGRU fits the behaviour well over a quite long period, although it appears to be slightly more damped than the ground truth. Moreover, the triangular-like waveshape induced by the nonlinear behavior is reproduced well by the FGRU. The results from FRNN fit the data during the initial cycle which has been seen during training, but then spurious damping and dispersion dominates. The reference FNO model is not able to fit the data at all, with training plateauing very early and never converging on a reasonable approximation. Again, this hierarchy seems to generalize to a wider set of examples, as is reflected in the validation MSE values seen above.

\subsubsection{2d Wave Equation}
\begin{figure}[h]
    \centering
        \includegraphics[width = 0.9 \columnwidth]{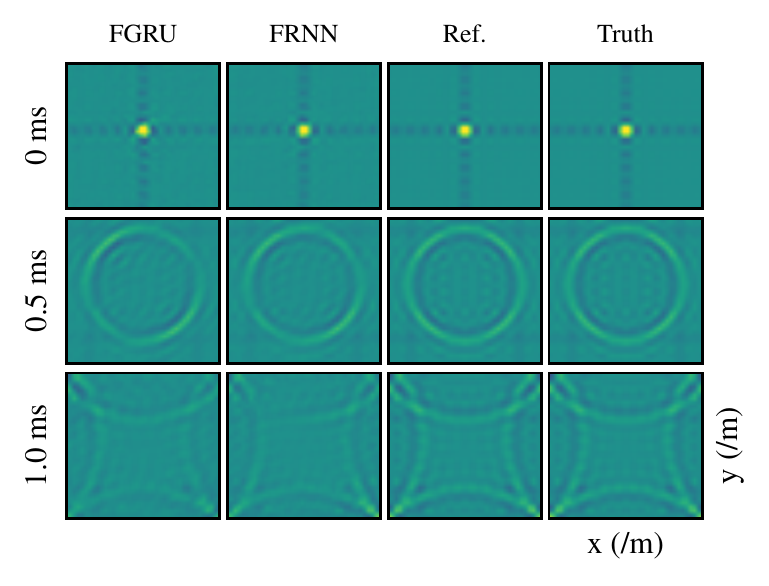}
    \caption{Spatial distribution of sound pressure at different points in time as obtained by the proposed FGRU and FRNN structures, the reference FNO model and the analytical solution obtained by the FTM model.}
    \label{fig:2d_wave_results}
\end{figure}

Fig.~\ref{fig:2d_wave_results} shows the result of exciting the trained models with an impulse halfway along the $x$ dimension, and slightly less than halfway along the $y$ dimension. In this case, the time period shown is around 2x that seen during training. In this case we see an interesting reversal of the results seen on the string models. The reference FNO model performs the best, with the FRNN and especially the FGRU falling off in accuracy after 1ms. These observations are in agreement with the validation MSE calculated over a wider range of examples. It is beyond the scope of the current work to properly examine why the hierarchy of performances is reversed in this case. An initial speculation might be that given the mathematically very simple structure of this model (despite its higher dimension count), the FNO might be benefiting from some kind of regularization effect gained by the constriction of the internal states back to the physical states between each time step. Further investigation of this phenomenon is left to future work.

\section{Conclusions}
\label{sec:conclusion}

In this work we gave an overview of existing methods of modeling PDEs using NNs. We presented two new structures based on the FNO approach, designed to perform better in audio-oriented tasks. We compared these structures theoretically to the FTM, and showed that they have significant mathematical parallels. This opens up an interesting new avenue in terms of the interpretability of these networks.

As an initial proof of concept, we trained the proposed network structures to reproduce datasets generated by existing physical models. The results of this training show that the proposed networks have potential for physical modeling in the audio domain. Future work could explore the applicability of such models to systems which are harder or more computationally expensive to model using existing techniques.

\bibliographystyle{IEEEbib}
\footnotesize
\bibliography{libdat} 

\end{document}